\documentclass[letterpaper]{sfchem}
\usepackage{graphicx}

\makeatletter
\newenvironment{tablehere*}
        {\def\@captype{table*}}
        {}


\def\cmd{cm$^{-2}$} 

\begin{document}

\title{Connection between PAHs and small hydrocarbons in the Horsehead Nebula Photo-Dissociation Region}

\author{David Teyssier\inst{1} \and David Foss\'e\inst{1} \and Maryvonne Gerin\inst{1} \and J\'er\^ome Pety\inst{1,2} \and Alain Abergel\inst{3} \and Emilie Habart\inst{4}} 
  \institute{LERMA, UMR 8540, CNRS/Observatoire de Paris, 24 rue Lhomond, 75231 Paris Cedex 5, France
	 \and IRAM, 300 rue de la Piscine, Domaine Universitaire, 38406 Saint Martin d'H\`eres, France
  \and IAS, UMR-8617, Universit\'e Paris-Sud, b\^atiment 121, 91405 Orsay, France
  \and Osservatorio Astrofisico di Arcetri, Largo E. Fermi 5, 50125 Florence, Italy} 
\authorrunning{Teyssier et al.}
\titlerunning{Connection between PAHs and small hydrocarbons in PDRs}

\maketitle 

\begin{abstract}
We present recent observations of small hydrocarbons (C$_3$H$_2$, C$_2$H, C$_4$H) with high abundances in the Photo-Dissociation Region of the Horsehead nebula. Our results show for the first time observational indications that the small hydrocarbon distribution follows the Aromatic Infrared Bands (AIBs) emission traced by ISO-LW2 (5-8.5\,$\mu$m), whereas it does not coincide with the CO and isotopes large-scale distribution. The derived abundances are significantly higher than in local clouds. This enhancement might be explained by an {\it in situ} formation assisted by the release of carbonaceous molecules from UV-irradiated aromatic particles.


\end{abstract}

\section{Introduction}
\label{intro}

The spatial distribution of the Aromatic Infrared Bands emission features (AIBs: 3.3, 6.2, 7.7, 8.6, 11.3 and 12.7 $\mu $m features) has been mapped at high spatial resolution by ISOCAM in many Photo-Dissociation Regions (PDRs). The identification of the AIBs carriers is still debated, however it is well established that they are complex carbon compounds heated transiently by the absorption of a single UV or visible photon. The best candidates are Polycyclic Aromatic Hydrocarbons (PAHs: e.g. L{\'e}ger \& Puget 1984) or Hydrogenated Amorphous Carbons (HACs: Duley \& Williams 1988). These large-sized carbonaceous molecules ($\sim$ 50 carbon atoms) are likely to be chemically connected to the numerous small hydrocarbon detected the ISM, including cyanopolyynes or acetylenic chains and rings. 

Various theoretical works and laboratory experiments support this point: modelling of dark cloud chemistry (Herbst 1991 ; McEwan et al. 1999), calculation of the acetylene formation rate by photo-erosion of UV irradiated PAHs (Allain et al. 1996), experiments of carbon chains production during the irradiation of HACs by a UV laser pulse (Scott et al. 1997). Nevertheless we still miss a firm observational link between AIB carriers and small hydrocarbons. PDRs are the best places to look for such a link as they exhibit strong AIB emission features which have been mapped at arcsec resolution by ISO in several objects. They are moreover highly structured regions where, due to the penetration depth of UV irradiation, the aromatic emission, the H$_2$ ro-vibrational lines, and the molecular (CO) emission are spatially separated (see Hollenbach \& Tielens 1997).

\section{Observations}
\label{observation}

\begin{figure*}
        \begin{center}
        \includegraphics[width=0.66\linewidth, angle=270]{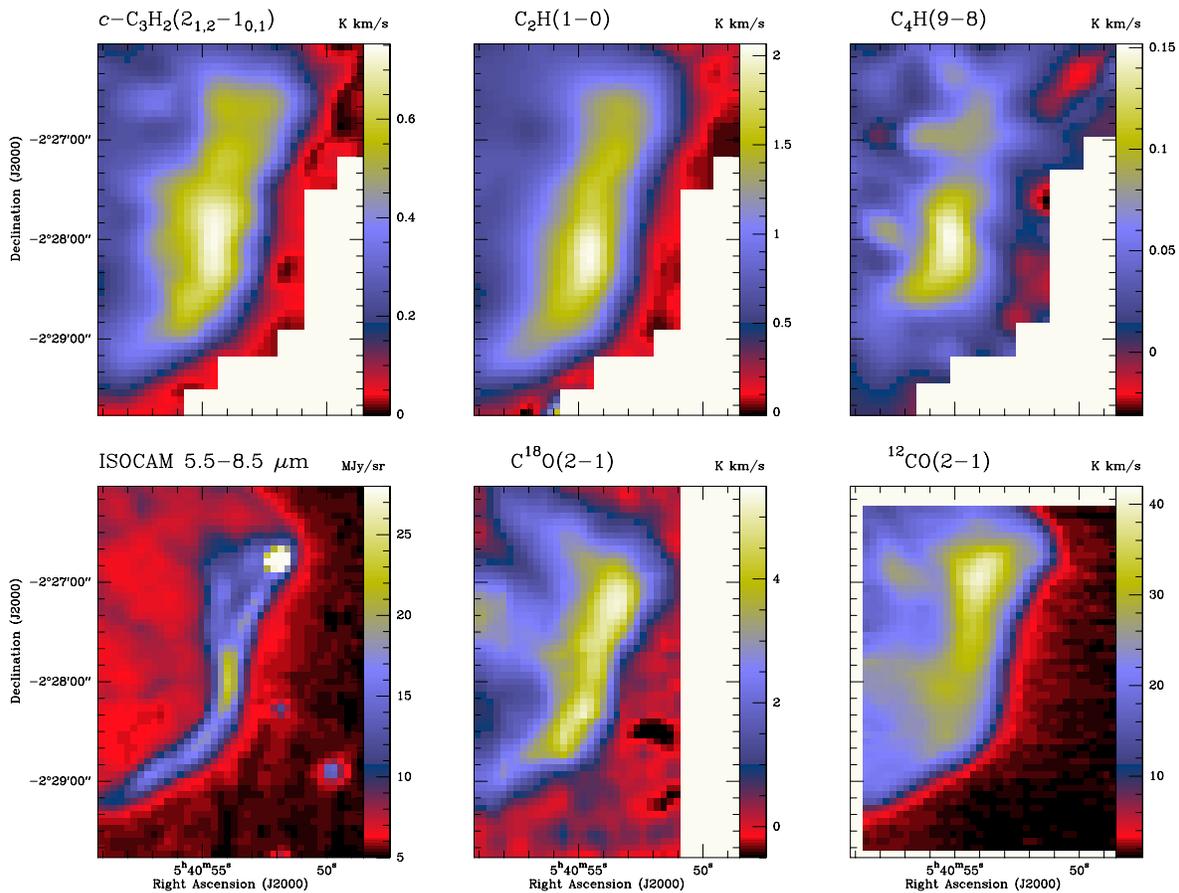}
        \caption{\label{hh fig poster}
Integrated maps ($T_A^*$, K\,km/s) of the molecular emission for various species, as well as the aromatic emission as covered by ISO.}
        \end{center}
\end{figure*}

We have used the IRAM 30-m telescope to observe small carbon chains (C$_2$H, $c$-C$_3$H, $l$-C$_3$H, $c$-C$_3$H$_2$, $l$-C$_3$H$_2$, and C$_4$H) in the PDR of the Horsehead nebula. We obtained maps of 2.5$^{\prime}\times$3.5$^{\prime}$ in three species. To our knowledge, these are the first maps of small hydrocarbons reported in photo-dissociation regions. In complement, we mapped with unprecedented spatial resolution the emission of the optically thick $^{12}$CO, and the likely optically thin C$^{18}$O isotope. These molecular data were compared to the AIB emission mapped by ISOCAM in the 5.5--8.5\,$\mu$m band (LW2 filter, Abergel et al. 2002). The considered data are compiled in Fig.~\ref{hh fig poster}.

Fig.~\ref{hh fig poster} shows that the small carbon chains are present until the edge of the PDR. The maps reveal moreover that all three hydrocarbon integrated intensities peak very close to the ISO-LW2 one, which is not true for the CO and isotope emission. There is perfect correlation between C$_2$H and $c$-C$_3$H$_2$, indicating likely similar physical conditions. This behavior holds for C$_4$H, but is still limited by a poor S/N ratio. A shift towards the molecular cloud is however observed, suggesting beam dilution of larger-scale structures. Interferometric data obtained at IRAM PdBI (Pety et al., in preparation) are no more affected by this effect and show a perfect coincidence of the hydrocarbon and aromatic emission peaks. At first sight, there seem to be a spatial correlation between aromatic particles and small hydrocarbons.

\section{Column densities}

We have computed the molecular column densities in the areas corresponding to the AIB and the CO peaks respectively. 
Tab.~\ref{tab coldens} gathers the results and shows that the hydrocarbon column densities are maximum at the aromatic emission peak, which is not the case for C$^{18}$O. The values are of the same order as those observed in the TMC-1 dark cloud, known as an efficient hydrocarbon factory. However the illumination conditions ($G_0=100$) are different from the dark cloud ones and small hydrocarbons are expected to be rapidly destroyed by the intense radiation fields.

\section{Discussion}

\begin{figure*}
        \begin{center}
        \includegraphics[width=0.8\linewidth]{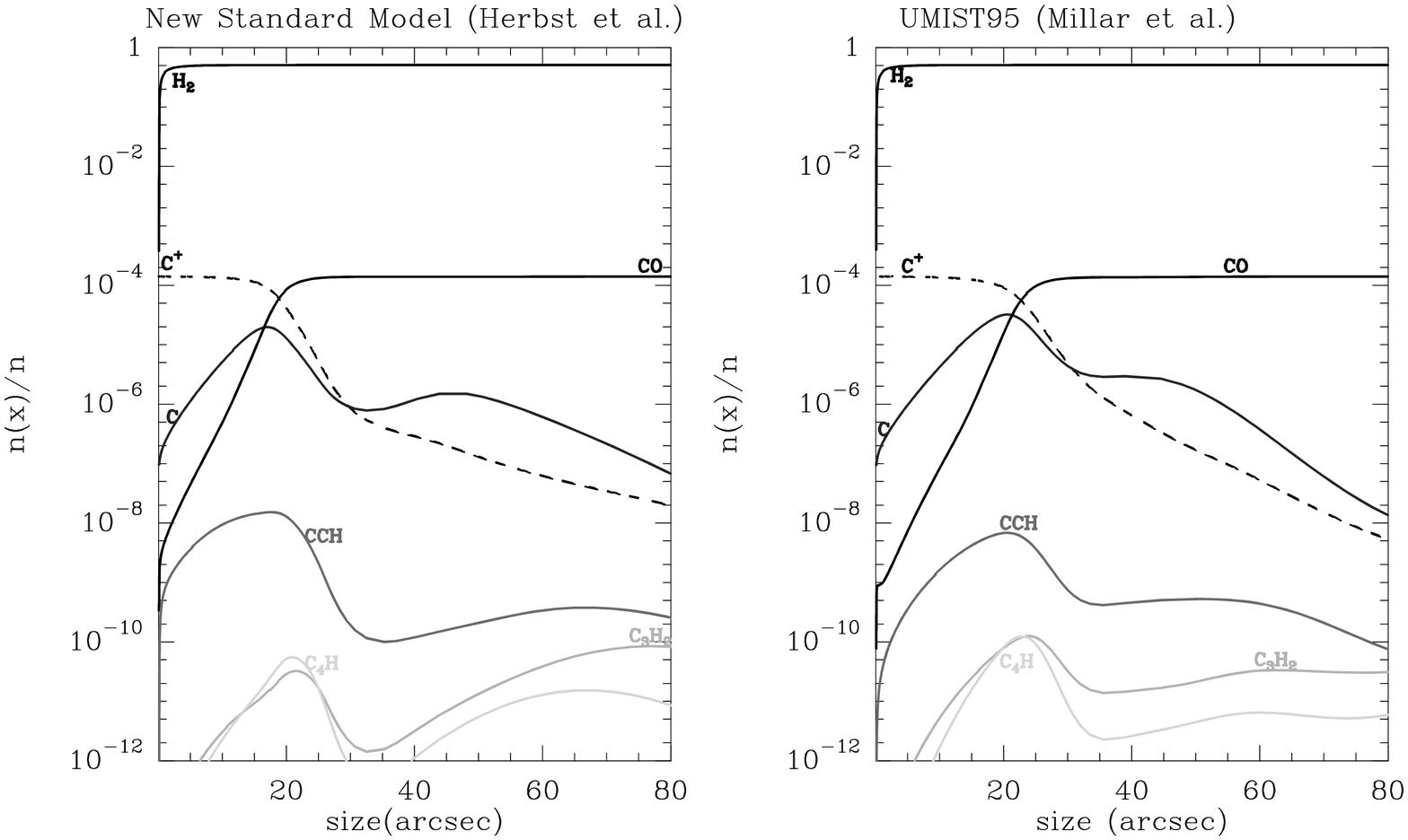}
        \caption{\label{models}
Abundances of various species derived from an updated Le Bourlot et al.'s PDR model, using two chemistry networks (calculations courtesy of E. Roueff).}
        \end{center}
\end{figure*}

\begin{table*}
\begin{center}
\begin{tabular}{l c c| c| c c}
\hline \hline
         &  Observed  &           &  \multicolumn{3}{c}{Column Density (\cmd)} \\ \cline{4-6}
Molecule &  Frequency & Position  &  Measured    &  \multicolumn{2}{|c}{Models} \\
         &  (GHz)      &           &              &  $A_V^{\rm peak} = 3$          &  $A_V^{\rm peak} = 15$  \\ \hline
C$_2$H   & 87.360   & AIB peak   &  1.8$\pm0.2$$\times$$10^{14}$ &  1.9--5$\times$$10^{13}$     &  1--2.5$\times$$10^{14}$      \\
         &          & CO peak   &  1.3$\pm0.1$$\times$$10^{14}$ &               &               \\
C$_3$H$_2$ & 85.339 & AIB peak   &  1-2$\times$$10^{13}$      &  1--3$\times$$10^{11}$  &  0.5--1.5$\times$$10^{12}$    \\
         &          & CO peak   &  0.7-1.4$\times$$10^{13}$  &                  &               \\
C$_4$H   & 85.634   & AIB peak   &  2$\times$$10^{13}$       &  1.7--3$\times$$10^{11}$ &  0.9--1.5$\times$$10^{12}$    \\
         &          & CO peak   &  1.3$\times$$10^{13}$      &                  &               \\
C$^{18}$O &         & AIB peak   &  4$\pm0.2$$\times$$10^{15}$&  5.4$\times$$10^{14}$   &  2.7$\times$$10^{15}$ \\
         &          & CO peak   &  5$\pm0.2$$\times$$10^{15}$&                  &               \\ \hline
\end{tabular}
\end{center}
\caption{\label{tab coldens}
Molecular column densities measured in the PDR (use of LTE and LVG approximations), and inferred from models (Fig.~\ref{models}) for two peak extinctions, in areas corresponding to the AIB and the CO emission peaks. Assumptions are: $n$(H$_2) = 10^4$\,cm$^{-3}$, $T_{\rm ex} \sim 10$\,K for all hydrocarbons, and $T_{\rm kin} \sim 35$\,K for C$^{18}$O.}
\end{table*}

We have compared our measurements to PDR models using several chemistry networks (Fig.~\ref{models}) and assuming representative conditions ($G_0=100$, $n_{\rm H}=2\times10^{4}$\,cm$^{-3}$). We find the following results:
\begin{itemize}
        \item the models reproduce well the abundance ratios of [C$_3$H$_2$]/[C$_4$H] and [C$^{18}$O]/[C$_2$H] ($\sim 1$ and 20 resp.) 
        \item  C$_3$H$_2$ and C$_4$H are at least one order of magnitude more abundant than expected compared to C$_2$H
\end{itemize}    

Part of the explanation possibly lies in the under-estimated H$_2$ column density. Kramer et al. (1996) reported $A_V=3$ inside the nebula (2$^{\prime}$ resolution), translating into molecular column densities 10-100 times lower than observed. Recent continuum observations performed at 1.2\,mm (11$^{\prime\prime}$ resolution) suggest visible extinctions more likely in the range $A_V=10-20$, which reconciles the observed C$^{18}$O and C$_2$H results with the model calculations. Nevertheless, a residual over-abundance (factor $\sim 20$) remains for the more complex hydrocarbons, suggesting that the hydrocarbon production is enhanced in the illuminated front. {\bf We propose that this formation process is related to the photo-erosion of UV-irradiated large carbonaceous compounds at the border of the PDR}. In this scenario, complex hydrocarbons such as C$_4$H would survive the intense illumination conditions encountered at the molecular cloud interfaces. This also suggests that heavier species (C4, C5) could be observed in UV-irradiated regions.




\end{document}